\documentclass[preprint,12pt]{elsarticle}

\usepackage{amssymb}
\usepackage{amsmath}
\usepackage{latexsym}
\usepackage{graphicx}
\journal{Computers \& Mathematics with Applications}

\begin{document}

\begin{frontmatter}



\title{Reserve Requirement Analysis using a Dynamical System of a Bank based on Monti-Klein model of Bank's Profit Function}

\author[label1,label2]{N. Sumarti}
\author[label3,label4]{I. Gunadi}
\address[label1]{Industrial and Financial Mathematics Research Group, Institut Teknologi Bandung, \\
Ganesha 10 Bandung, Indonesia
}
\address[label3]{Directorate of Economic Research and Monetary Policy, Bank of Indonesia, \\ Jl. M.H. Thamrin no. 2, Jakarta, Indonesia}
\address[label2]{novriana@math.itb.ac.id}
\address[label4]{imangunadi@bi.go.id}
%% \author{}
%% \address{}

\begin{abstract}
Commercial banks and other depository institutions in some countries are required to hold in reserve against deposits made by their customers at their Central Bank or Federal Reserve. Although some countries have been eliminated
it, this requirement is useful as one of many Central Bank's regulation made to control rate of inflation and conditions of excess liquidity in banks which could affect the monetary stability.  The amount of this reserve is affected by the volumes of the commercial
bank's loan and deposit, and also by the bank's Loan to Deposit Ratio (LDR) value. In this research, a dynamical system of the volume of deposits (dD/dt) and loans (dL/dt) of a bank is constructed from the bank profit equation
by Monti-Klein. The model is implemented using the regulation of Bank of
Indonesia, and analysed in terms of the behaviour of the solution. Based on some simplifying assumptions in this model, the results show that eventhough the LDR values at the initial points of two solutions are the same, the behavior of solutions will be significantly different due to different magnitude of L and D volumes.
\end{abstract}

\begin{keyword}
Reserve requirement, model of bank, dynamical system, numerical method
%% or \MSC[2008]  \sep code (2000 is the default)
\MSC 34N05  \sep 37N40
\end{keyword}

\end{frontmatter}

% \linenumbers

\section{Introduction}
\label{lab1}
The need of reliable techniques to model the behaviour of comercial banks
in a country
in a period of time is inevitable nowdays. The model could benefit in the decisions
for the appropriate regulation of banking industries be applied. Recently the derivation of dynamical models for macroeconomics is more in
quantitative approach rather than modelling approach such as using dynamical
systems. Some papers examining stochastics different equations
for interest rate models are including \cite{[1a],[1b],[1c]}.  In \cite{[1d]},
a diffusive Lotka–Volterra system was formulated that represents the dynamics of market share with the implementation on data from Pakistan. In \cite{[2]},
dynamical
models on bank profit was built to analysed the Return-on-Assets (RoA) and Return-on-Equity (RoE) of a bank. In \cite{[3]}, the economic aspects of the stochatic dynamics model of bank assets and and liabilities was discussed.\\

In \cite{[0a]}, the question of the interdependence between
decisions about loans and deposits in the banking regulation is answered
using the Monti-Klein model. In \cite{[0b]}, the authors explained the scopes
of economies between loan and deposit in an oligopolistic market based on
the same model. \\

A new model proposed in this research in order to achieve a convenient form of the model to show the interdepence
of loan and deposit via a system of dynamical model. The model is to forecast
the volumes of loan and deposit of commercial banks in a general period of time. A dynamical model consists
of a set of differential equations, that can determine the observed states, for example bank's loan and deposit volumes, for all future times based on the current state. It can be shown in its phase portrait where the behaviour of the solution's
trajectories and velocities are depicted in all conditions. It will also could show small changes in the state of the system create either small or big changes in the future depending on the model. In its stability analysis on a dynamical system, the equilibrium points whereabout the velocity of the solutions is zero, are discovered in order to understand the behaviour of their nearby solutions. For instance, if they are going in spiral or star-like trajectories
approaching these points, then it is the stable condition. Or they are passing by or going out of these points, then it is the unstable condition. These
behaviours can be interpreted into the conditions of the bank in general.
From the obtained dynamical model, we can do the stability analysis of its equilibrium points. Detailed explanation of this analysis can be found in dynamical system textbooks, which are including \cite{[4],[4b]}. \\

The dynamical model constructed in this paper is implemented for calculating the reserve requirement
regulated by the Central Bank, who is responsible for achieving and maintaining monetary stability. This regulation is made to control rate of inflation and conditions of excess liquidity in banks. However, the requirement imposes a cost on the private sector equal to the amount of forgone interest on these reserves \cite{[4c]}. It is said that the higher the level of reserve requirements, the greater the costs imposed on the private sector. At the same time, however, higher reserve requirements may smooth the implementation of monetary policy and damp volatility in the reserves market. In \cite{[5]}, the introduction of the reserves
gives impact on the existence and efficiency properties of Nash equilibria
of the model of double Bertrand competition. In \cite{[5a]}, the estimation of optimal reserve holdings for countries under various monetary regimes
was proposed. In \cite{[6]}, the significance of interest rates on the optimal
international reserve holdings was discussed.\\ 

In Indonesia, there are primary
reserves that give no interest and secondary reserves that give interest.
We will use Bank of Indonesia regulation in \cite{[1]} in order to see the
total amount of these reserves.  One of types of the reserves needs the ratio of Loan and Deposit volumes. With the dynamical system
derived in this paper, we can describe the fluctuation of the values of reserve requirement from a bank in Indonesia.
The model can be easily adapt to other countries' regulations.


\section{Mathematical Model of Banking}
\label{lab2}
Having improved Monti-Klein model \cite{[7],[8]}, a bank's profit function can be expressed by the optimization problem below:
\begin{equation} \mbox{max } \ \pi = r_L L+r M+r_B B + r_{R_2} R_2- r_D D-C(D,L) \label{(1)}\end{equation}
\begin{equation} \mbox{such that } L+M+R_1+R_2+B= D+K   
 \label{(2)}\end{equation}
where $L$ and $D$ are volumes of loan and deposit. $M, R_1, R_2, K$ and $B$ are respectively volumes of net position of the bank on the interbank market, the primary and secondary reserves, the amount of equity held by the bank and government Treasury bills. $r_L,r_D,r_M,r_{R_2}$ and $r_B$ are
respectively the related interest rates of $L, D, M, R_2$ and $B$. Note that we assume there is no interest rate for $R_1$ and $K$. $C(D,L)$ is a cost function that describes the bank's management costs of servicing loans $L$ and deposits $D$.\\ 

Based on \cite{[1]}, the reserve requirement or GWM (\textit{Giro Wajib Minimum}) in
Indonesian Language are clasified into primary $R_1$, secondary $R_2$ and LDR (Loan to Deposit Ratio). $R_1$ and  $R_2$ are proportion of the volume of the deposits, or $R_1=\kappa_1 D$ and  $R_2=\kappa_2 D$.  The LDR GWM will be explained in detail in the next section. Stated in the previous paragraph, only the secondary GWM could bear interest with rate $r_{R_2}$. $B$ and $K$ are proportions of the volumes of the deposit and loan respectively, or  $B=\delta D$ and
$K=\gamma L$. The net position of the bank on interbank market is defined as
\begin{equation} M=(1-\kappa_1-\kappa_2-\delta)D+L(\gamma-1).\label{(3)}
\end{equation} 
The profit function in equation (\ref{(1)}) becomes
\begin{equation}\begin{split}
\pi=(r(1-\kappa_1-\kappa_2-\delta)+r_B \delta+r_{R_2} \kappa_2-r_D )D +\\
        (r_L+r\gamma-r)L-C(D,L).     \label{(4)}
\end{split}
\end{equation} 
For simplicity, some assumptions are applied:
\begin{enumerate}
\item No stochastic change on rate of return of deposit and the bond. 
\item No risk of default on loan.
\item No liquidity problem occurs.
\item No various types of loan and deposit, so they are unique.
\item The cost function has been simplified.
\end{enumerate}
In \cite{[2]}, the monopoly bank which represents the banking industry as a whole, have a downward-sloping demand for loans with respect to the loan rate and an upward-sloping demand for deposits with respect to the deposit rate. Later, those two conditions will be a validation later whether the model is acceptible or not.\\

A dynamical system of $D$ and $L$ is obtained by using chain rule in derivation on these volumes with respect to time, 
\[ \frac{dD}{dt}=\frac{\partial D}{\partial \pi}\frac{\partial \pi}{\partial r_D} \frac{dr_D}{dt} \mbox{ and  } \frac{dL}{dt}=\frac{\partial L}{\partial \pi}\frac{\partial \pi}{\partial r_L} \frac{dr_L}{dt}\]
where $\frac{\partial \pi}{\partial r_D}$ and $\frac{\partial \pi}{\partial r_L}$ are respectively from the partial derivations of the profit function $\pi$  in \label{(6)} with respect to $r_D$ and  $r_L$. In the similar way, we also have $\frac{\partial D}{\partial \pi}$ and $\frac{\partial L}{\partial \pi}$
from $(\frac{\partial D}{\partial \pi})^{-1}$ and $(\frac{\partial L}{\partial \pi})^{-1}$ respectively. Now we need to define $\frac{dr_D}{dt}$ and $\frac{dr_L}{dt}$. Having in the first stage of the mathematical modeling, we define the loan
and deposit's interest rates using the simplification assumption that these quantities are varying periodically as in functions of cosine and sine. In the further stages, we can define those functions which are more related to the real world, so more complicated analysis will be needed. The constructed
dynamical system of $D$ and $L$ will be
\begin{equation}\label{(5)}
\begin{array}{lcl}
\frac{dD}{dt} &=& \frac{\partial D}{\partial r_D}\frac{d r_D}{dt}-\frac{D
\frac{d r_D}{dt}}{r(1-\kappa_1-\kappa_2-\delta)+r_B \delta -r_{R_2} \kappa_2
- r_D \frac{\partial C}{\partial D}}\\
\frac{dL}{dt} &=& \frac{\partial D}{\partial r_D}\frac{d r_D}{dt}-\frac{L
\frac{d r_L}{dt}}{r_L + r \gamma -r-\frac{\partial C}{\partial D}}
\end{array}
\end{equation}
Now we assume that $\frac{\partial C}{\partial D}=\frac{\partial C}{\partial L}=c_1 D+c_2 L$, so
\[ C(D,L)=kDL+\frac{1}{2} kL^2+\frac{1}{2} kD^2,\]
where $c_1,c_2$ and $k$ are constants. Other assumptions are needed to be able to find the solution of the system (\ref{(5)}) numerically. We assume
$\frac{\partial D}{\partial r_D}=b, \ 0<b\leq 1$, and $\frac{\partial L}{\partial r_L}=g, \ -1\leq g< 0$, which can be intrepretated as the public's respond to the alteration of the interest rates of loan and deposit. The values of $r_D$,$r_L$ and $r$, which are given as the approximate situation in Indonesia, range respectively from $0.02$ to $0.06$, $0.08$ to $0.14$, and $0.06$ to $0.07$. To model the fluctuation of these values, we assume that 
\begin{equation}\begin{array}{l}
r_D=0.04+0.0\sin{(2\pi t)},\\
r_L=0.11+0.03\cos{(2\pi t)},\\
r =0.06+0.01 \sin{(2\pi t)}, 
\end{array}
\end{equation}
Having substituting some values of parameters, the numerical solution can be obtained and analysis involving equilibrium points of the dynamical system will be discussed in \cite{[9]}.
In the next section we will discus the analysis of the behaviour of the solutions numerically on this system and the analysis of GWM values based on the different values of the deposits and loans demands.
  
\section{Reserve requirement model in Indonesia}
The reserve requirement (or GWM) in Rupiah, Indonesian currency, was stipulated in the Regulations  of Bank Indonesia \cite{[1]}. We need to make mathematical expressions of the regulation. The proportion of the
primer GWM is $8\%$ of the banks'deposit volume, or $\kappa_1=0.08$, and
of the secondary GWM is $2.5\%$ or $\kappa_2=0.025$.\\ 

Loan to Deposit Ratio (LDR) is the ratio of  the volume of loan given to the third party in the Rupiah and foreign currencies, excluding the loan for other bank,  with respect to the volume of deposit given to the third party including account balances, deposits in Rupiah and other currencies, excluding deposit from the interbank market.  Here LDR function $\lambda(t)$ is defined as follows 
\[\lambda(t)= \frac{L(t)}{D(t)}.\]
Based on \cite{[1]}, the calculation of LDR needs definition of some parameters. There are the lower and upper LDR-targets which are $\lambda_l=78\%$ and $\lambda_u=100\%$ respectively. Upper disincentive parameter  $\gamma_u=0.2$
 is a multiplication factor used in the calculation when the Bank has LDR value higher than the upper target $\lambda_u$. Similarly, the lower disincentive parameter $\gamma_l=0.1$ is used when the bank has LDR value smaller than
the lower target $\lambda_l$.  See equation (\ref{(7)}) where those apameters
are applied. Capital Adequacy Ratio (CAR) or $\mu(t)$ is the ratio of bank's capital to its risk and it is used to watch the health condition of the bank. Denote $\mu_m$ the minimum CAR. The reserve function GWM$(t)$ is defined as follows:
\begin{equation}\label{(7)}
\begin{array}{ccl}
\mbox{GWM}(t)&=&
\left\{ 
\begin{array}{cl}
\gamma_l \Delta_l \lambda(t) D(t) & \mbox{ if } \lambda(t) < \lambda_l,\\
0 & \mbox{ if } \lambda_l \leq \lambda(t) \leq \lambda_u,\\
\gamma_u \Delta_u \lambda(t) D(t) & \mbox{ if } \lambda(t) > \lambda_u, \mbox{
and } \mu(t) < \mu_m,\\
0 & \mbox{ if } \lambda(t) > \lambda_u,\mbox{ and } \mu(t) \geq \mu_m,\\
\end{array}
\right.
\end{array}
\end{equation}
Here $\Delta_l \lambda(t)=\lambda_l-\lambda(t)$ and $\Delta_u \lambda(t)=\lambda(t)-\lambda_u$. In this paper, we assume that $\mu(t)<\mu_m$ for all $t$. 

\section{Numerical Solution}
\begin{figure}[htb]
\centering\noindent
\includegraphics[width=2in]{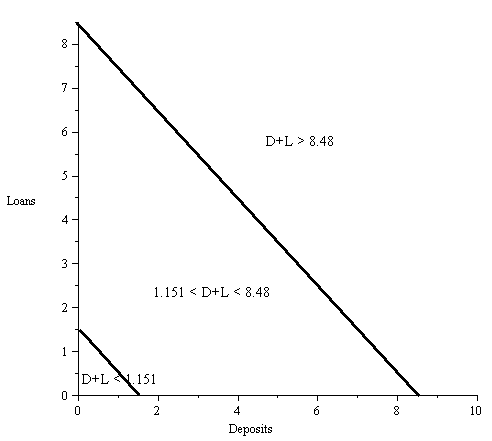}
\caption{Three areas for initial values}\label{figure-1}
\end{figure}

In this section we simulate the dynamical system (\ref{(5)}). 
The resulted dynamical system will have singular values of $D(t)$ and $L(t)$
for all $t$  where the system will not defined. Those values are satisfied below equations. 
\begin{equation}\label{(10)}
\begin{array}{l}
0.01 (D(t)+L(t))=1.515\times 10^{-2}-1.145 \times 10^{-2} \sin{(2\pi t)}\\
0.01 (D(t)+L(t))=5.48\times 10^{-2}+3 \times 10^{-2} \cos{(2\pi t)}-9.2\times
10^{-3} \sin{(2\pi t)}.
\end{array}
\end{equation}
When $t=0$, the above equations give 3 regions for the initial points (see
Figure \ref{figure-1}) to be chosen for solving the dynamical system numerically.
The initial values discussed in this paper are chosen with particular values of $L(t)$ and $D(t)$ ratio, those are values from $0.2$ to $2$. In the figures
presented in this section, solutions to the system with those initial values are named alphabetically as A to J. The initial values of $L(t)$ and $D(t)$ that show their volumes are roughly chosen each from the above areas, so there are 3 sets of initial values with the same ratios that will be analysed. The first set have constant $D(0) = 0.7$, the second set have constant $D(0) = 6$, and the last set have constant $D(0) = 10$.  It will be shown that the obtained results have different behaviour of solutions with respect to these different sets of initial values.\\
\begin{figure}[htb]
\centering\noindent
\includegraphics[width=2in]{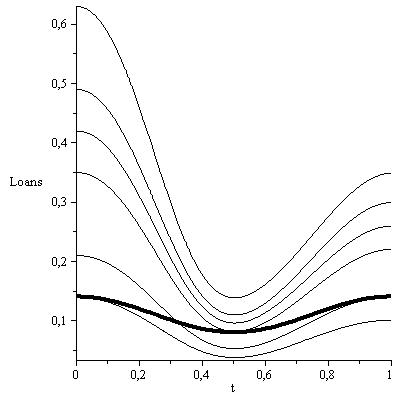}
\includegraphics[width=2in]{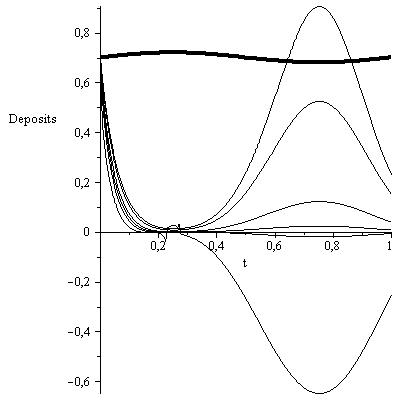}
\caption{Plots of $L(t)$ (left) and $D(t)$ (right) compared with their interest
rates (in bold curves) for initial points of Set 1.}\label{figure-2}
\end{figure}
\begin{figure}[htb]
\centering\noindent
\includegraphics[width=2in]{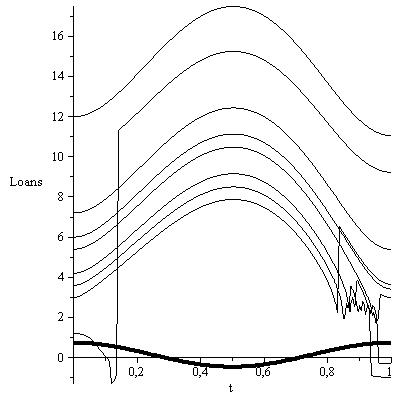}
\includegraphics[width=2in]{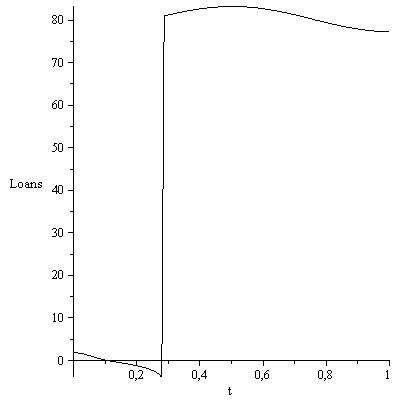}
\caption{$L(t)$ plots for A, C, D, E, F, G, H, I(left) and B (right) are
compared
to their interest rate plot (bold curve in the left graph) for initial points of Set 2.}\label{figure-3}
\end{figure}

Based on \cite{[2]} where there will be downward-sloping demand for loans with respect to the loan rate and an upward-sloping demand for deposits with respect to the deposit rate, the numerical solution obtained for deposit and loan from the first set does not follow this behaviour, as seen in Figure \ref{figure-2}. The graphs of loan decrease and increase consistently with the loan interest rate. This behaviour contradicts with the fact that the increase of loan interest rate will discourage people to take loan so its \begin{figure}[htb]
\centering\noindent
\includegraphics[width=2in]{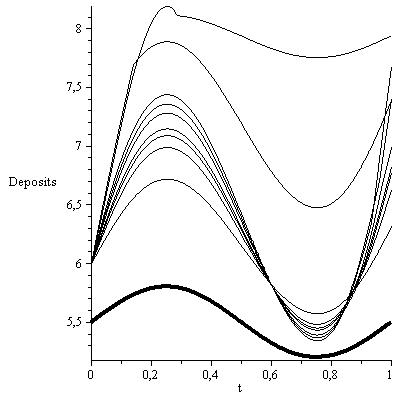}
\caption{Plots of $D(t)$ and their interest rate (
bold curve ) for initial points of Set 2.}\label{figure-4}
\end{figure}
\begin{figure}[htb]
\centering\noindent
\includegraphics[width=2in]{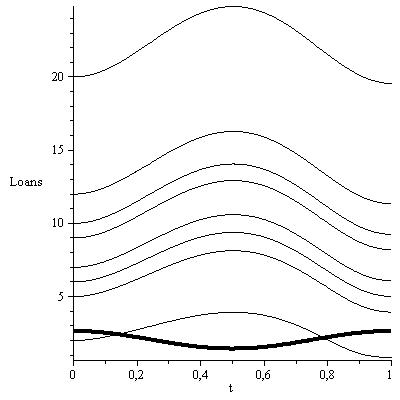}
\includegraphics[width=2in]{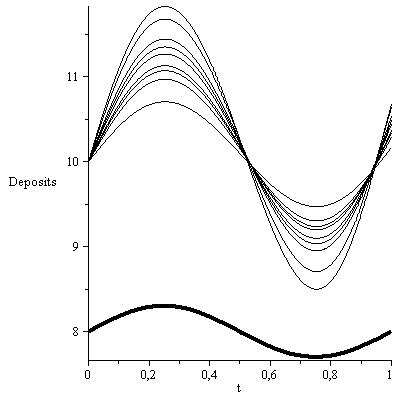}
\caption{Plots of $L(t)$(left) and $D(t)$ (right) are compared to their interest rates (bold curves) for initial points of Set 3.}\label{figure-5}
\end{figure}
volume will decrease. Furthermore, the graphs of deposit volumes show inconsistence when we compare it to the behaviour of the deposit interest rate in figure
\ref{figure-2}. This is also incorrect behaviour because they should be consistent. Due to these results, we will not use solutions from the first set in calculating the LDR reserves. Having obtained solutions from the second set in Figure \ref{figure-3}, we observed that the behaviour of the loan graphs is opposite to the behaviour of the loan interest rate, except for B where $(D(0),L(0)) = (6, 1.8)$. Some singular values are shown in the beginning of solution A, and at the end parts of solutions B, C, D. In Figure \ref{figure-4}, the behaviour of the deposit graphs is consistent with the behaviour of the deposit interest rate. The graph of solution B at the upper curve shows a different behaviour with other solutions. 
The results from the second set is satisfactory, except for solution B, so we will use some of the solution in calculating the LDR reserves. For solutions from the third set, the behavior of loan and deposit volumes and their interest rates are all correct, which are shown in Figure \ref{figure-5}. Solutions from this set will be used.\\

\begin{figure}[htb]
\centering\noindent

\includegraphics[width=2in]{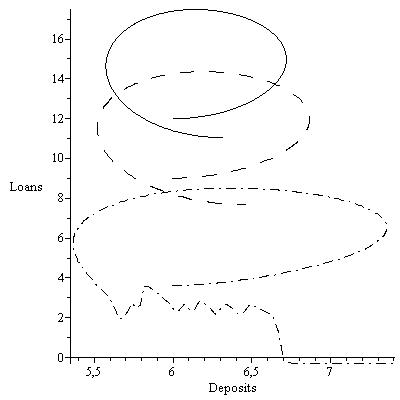}
\includegraphics[width=2in]{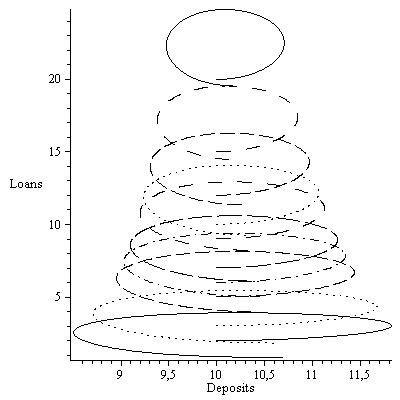}
\caption{Plots of $L(t)$ with respect to $D(t)$ Set 2 (left) and Set 3 (right).}\label{figure-6}
\end{figure}

Now we plot $L(t)$ with respect to $D(t)$ in Figure \ref{figure-6}. It shows that the graphs of solutions for some initial points from Sets 2 on the left and 3 on the righ. In the cartesian axes, the graphs of solutions with the smaller initial points are in the same time under the graphs of solutions from the higher initial points.
\begin{figure}[htb]
\centering\noindent
\includegraphics[width=2in]{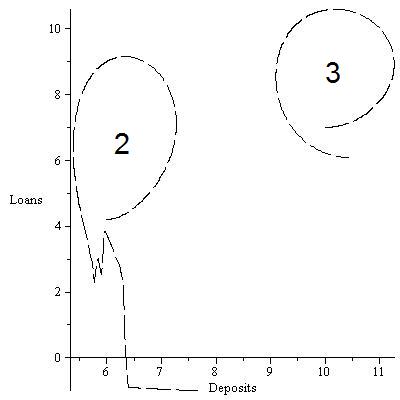}
\includegraphics[width=2in]{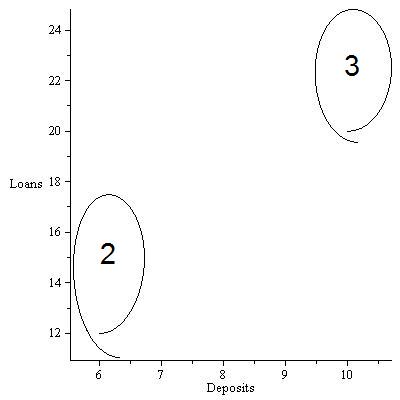}
\caption{Comparison of $L(t)$ and $D(t)$ plots for point E (left) and J (right).}\label{figure-7}
\end{figure}
Figure \ref{figure-7} shows the solution's graphs for the same initial points but from different set. The left is for plots of solution E and the right
is for plots of solution J. Both plots of solution E are similar but solution from set 2 has singular values at the end of plot. The ellipse-like plot from set 2 has a bit more flattened than plot from set 3. Although their
LDR values are the same, the magnitude of solutions 2 is lower than solutions
3. This also occurs in for solution J on the right side of Figure \ref{figure-7}.\\
\begin{figure}[htb]
\centering\noindent
\includegraphics[width=2in]{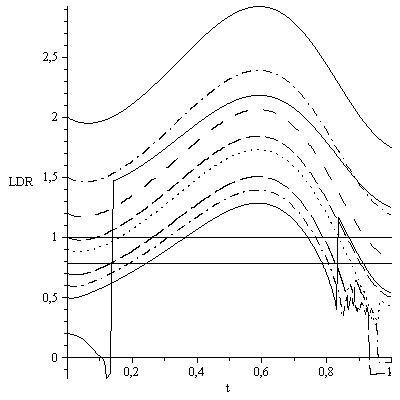}
\includegraphics[width=2in]{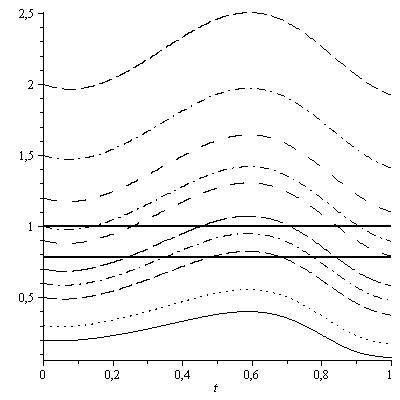}
\caption{Plots of $L(t)$ and $D(t)$ ratios for Set 2 (left) and Set 3 (right).}\label{figure-8}
\end{figure}
\begin{figure}[htb]
\centering\noindent
\includegraphics[width=2in]{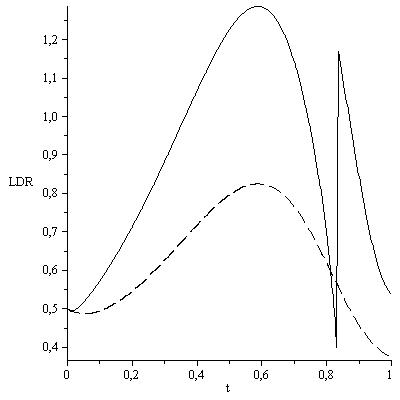}
\includegraphics[width=2in]{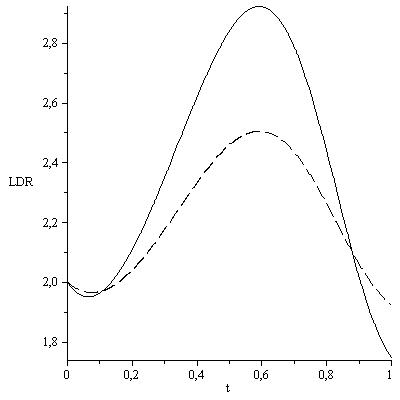}
\caption{Comparison of LDR plots for Set 2 (dash) and Set 3 (solid).}\label{figure-9}
\end{figure}

\section{Reserve Requirement Analysis}
Now we implement the obtained solutions to the dynamical model in order to
analyse the behaviour of the reserve or GWM if it is calculated using
Loan to Deposit Ratio (LDR). Figure \ref{figure-8} shows plots of $L(t)$ and $D(t)$ ratio for set 2 (left) and 3 (right). Remmeber that these
sets yield solutions from the same ratios with respet to different values
of $D(0)$. Here $D(0)=6$ for set 2 and $D(0)=10$ for set 3. Two bold lines in both graphs are the condition when values of LDR are $\lambda_l=78\%$ and $\lambda_u=100\%$, which will be used in the calculation of GWM. Note that instabilities occur
in solutions A and B of this set. It shows that solutions from smaller initial points are plotted under solutions of higher initial points. There is no solution from set 2 whose LDR
under $\lambda_l$ for whole period. On the other hand, there are 3 solutions
from set 3 whose LDR under $\lambda_l$.\\

There are significant differences of LDR values between points from set 2 and 3. Figure \ref{figure-9} shows the graph of solutions for C (left) and G (right). The LDR curves of set 2 (dash line) are plotted lower than of set 3 (solid line). It shows that the results from the same LDR at the initial point give significantly different magnitude. \\
\begin{figure}[htb]
\centering\noindent
\includegraphics[width=2in]{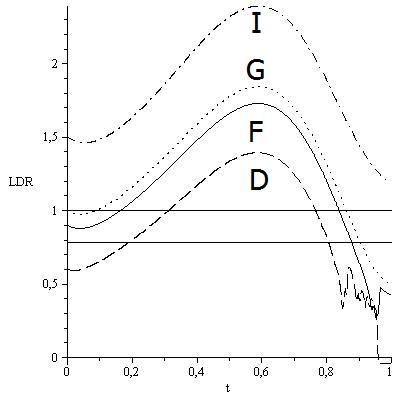}
\includegraphics[width=2in]{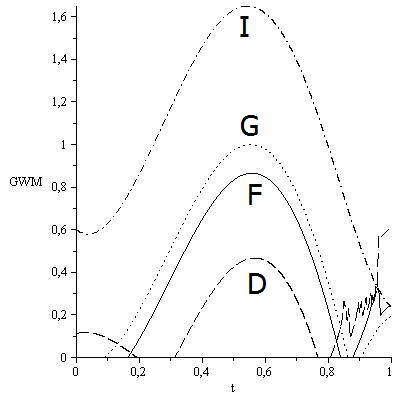}
\caption{LDR and GWM plots for Set 2: D(dash), F (solid), G (dot), and I (dashdot).}\label{figure-10}
\end{figure}
\begin{figure}[htb]
\centering\noindent
\includegraphics[width=2in]{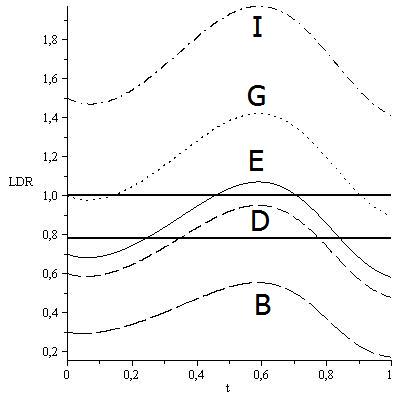}
\includegraphics[width=2in]{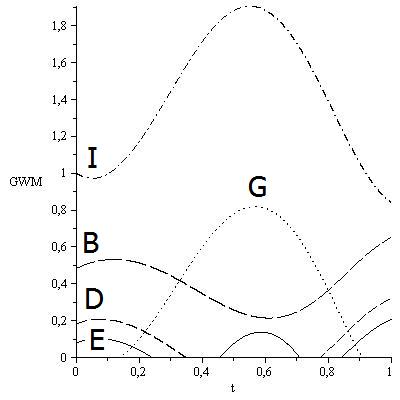}
\caption{LDR and GWM plots for Set 3: B(longdash), D (dash), E (solid), G (dot), and I (dashdot).}\label{figure-11}
\end{figure}

Now we calculate the GWM values for Set 2 and 3. Figure \ref{figure-10} right shows the graph of GWM values of D, F, G and I from Set 2. If we observe Figure \ref{figure-10} left, the LDR graph of $D$ is in regions under $78\%$, between $78\%$ and $100\%$, and above $100\%$. Its GWM values where
$78\% \le\lambda_l \le 100\%$ are consistently $0$ in two intervals. The LDR graphs of F and G give values are also in 3 regions. The upper curve is I where it only gives area above $100\%$. It seems in general that the higher ratio of L and D, the higher value of GWM. It also shows the larger
area under the GWM curve. \\

The graphs of GWM values for Set 3 is shown in Figure \ref{figure-11} right
which is related to Figure \ref{figure-8} right, which is plotted again for some
particular solutions in Figure \ref{figure-11} left. In that figure, solutions A and B are in LDR region only under $78\%$. Solutions C - D and F - G are in  2 regions, solution E is in 3 regions, and solutions H, I, J are in one region which is only above $100\%$. It gives different behaviour where the higher ratio of L and D is not always the higher value of GWM, except for
solutions H-J. Note that H to J are solutions with LDR between $1.2$ to $2$.\\ 
\begin{figure}[htb]
\centering\noindent
\includegraphics[width=2in]{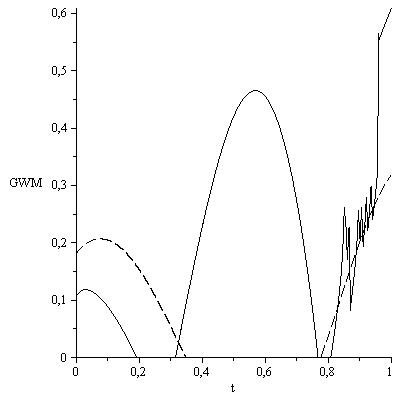}
\includegraphics[width=2in]{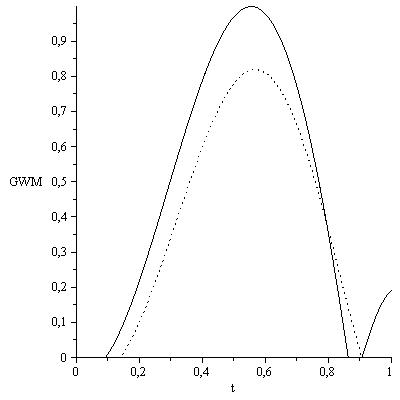}\\
\includegraphics[width=2in]{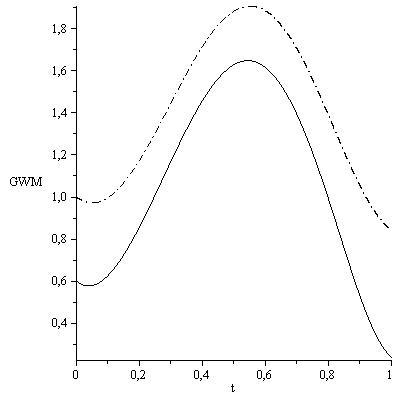}
\caption{Comparison of some GWM plots for Set 2 (solid lines) and Set 3 (others):
D (upper left), G (upper right) and I (lower).}\label{figure-12}
\end{figure}

Figure \ref{figure-12} shows the comparison of the GWM plots of solutions
D, G and I from Set 2 (solid lines) and Set 3. There are significantly different between the plots from those sets. The areas under the curves of solutions D and G from Set 2  are smaller than the related areas from Set 3. It shows
that larger amount of GWM collected in a particular time is resulted from
the bank which has higher initial value of LDR. However,
 the areas under solution I from Set 2 is larger than of from Set 3, which
shows not the same behaviour as the previous.

\section{Conclusions}
The dynamical system derived from the bank profit function can describe the properties of the deposit and loan functions, and the types of reserve requirement functions occured.   There are significant differences on the behaviour of solutions from Set 1, 2, and 3, which have the same ratio of L and D at the initial points. Those are shown in the plots of L to D, LDR and GWM values.
Only for set 2, the amount of GWM collected in a period of time will increase when the LDR ratio increases. The latter behaviour applies to solutions from
set 3 when the LDR values are between 1.2 to 2. For further research, we suggest to change different values of paramaters in the GWM function based on the magnitude of the L and D volumes in order to control the desired amount
of the GWM function in a particular period.

\section*{Acknowlegement}
We are very grateful for valuable inputs in this research from M.R.E. Proctor from University of Cambridge and J.S. Chadha from University of Kent. This research is sponsored by 2012 Staff Exchange Program, IMHERE B.2C FMIPA ITB.
%% References


\bibliographystyle{model1a-num-names}
%%\bibliography{<your-bib-database>}



\end{document}